\documentclass[11pt, a4paper]{article}
\usepackage{CJKutf8}
\usepackage{geometry}
\usepackage{subfigure}

\usepackage{amssymb}
\usepackage{amsmath}
\usepackage{amsthm}
\usepackage{amsfonts}
\usepackage{mathtools}
\usepackage{mathrsfs}
\usepackage{bm} % 处理数学公式中的黑斜体的宏包

\usepackage{graphicx}
\usepackage{dcolumn}
\usepackage[ruled,lined,algonl]{algorithm2e}
\usepackage{color}
\usepackage{autobreak}
\allowdisplaybreaks

\theoremstyle{plain}
\newtheorem{theorem}{Theorem}

\newtheorem{proposition}{Proposition}

\theoremstyle{definition}

\newtheorem{model}{Model}

\theoremstyle{remark}

\DeclareMathOperator{\Det}{Det}
\DeclareMathOperator{\Tr}{Tr}
\DeclareMathOperator{\res}{res}
\DeclareMathOperator{\discr}{discr}
\usepackage{authblk}
\usepackage{listings}

\begin{document}
\begin{CJK}{UTF8}{gbsn}

\title{Stability of Cournot duopoly games with isoelastic demands and quadratic costs}

%\author{Xiaoliang Li\thanks{Corresponding author: xiaoliangbuaa@gmail.com}}
%\affil{School of Finance and Trade, Dongguan City College, Dongguan, P. R. China}

\author[a]{Xiaoliang Li}

\author[b]{Li Su\thanks{Corresponding author: suli0917@ruc.edu.cn}}

\affil[a]{School of Finance and Trade, Dongguan City College, Dongguan, China, 523419}

\affil[b]{School of Applied Economics, Renmin University of China, Beijing, China, 100872}

\maketitle

\begin{abstract}
In this discussion draft, we explore different duopoly games of players with quadratic costs, where the market is supposed to have the isoelastic demand. Different from the usual approaches based on numerical computations, the methods used in the present work are built on symbolic computations, which can produce analytical and rigorous results. Our investigations show that the stability regions are enlarged for the games considered in this work compared to their counterparts with linear costs, which generalizes the classical results of ``F. M. Fisher. The stability of the Cournot oligopoly solution: The effects of speeds of adjustment and increasing marginal costs. The Review of Economic Studies, 28(2):125--135, 1961.''.
\end{abstract}

%\section{Introduction}
%
%\cite{Agiza1998E,Agliari2000T,Agliari2006H,Ahmed1998D,Askar2016N,Bischi2007O,Cavalli2015Na,Fisher1961T,Puu1991C,Tramontana2010H}

\section{Models}

Let us consider a market served by two firms producing homogeneous products. We use $q_i(t)$ to denote the output of firm $i$ at period $t$. Moreover, the cost function of firm $i$ is supposed to be quadratic, i.e., $C_i(q_i)=c_iq_i^2$ with $c_i>0$. At each period $t$, firm $i$ first estimates the possible price $p_i^e(t)$ of the product, then the expected profit of firm $i$ would be 
$$\Pi_i^e(t)=p_i^e(t) q_i(t) - c_iq_i^2(t),~~i=1,2.$$
In order to maximize the expected profit,  at period $t$ each firm would decide the quantity of the output by solving
$$q_i(t)=\arg\max_{q_i(t)}\Pi_i^e(t)=\arg\max_{q_i(t)}\left[p_i^e(t) q_i(t) - c_iq_i^2(t)\right],~~i=1,2.$$

Furthermore, assume that the demand function of the market is isoelastic, which is founded on the hypothesis that the consumers have the Cobb-Douglas utility function. Hence, the real (not expected) price of the product should be
$$p(Q)=\frac{1}{Q}=\frac{1}{q_1+q_2},$$
where $Q=q_1+q_2$ is the total supply. Four types of players with distinct rationality degrees are involved in this draft, which are described as follows one by one.

A \emph{rational player} not only knows clearly the form of the price function, but also has complete information of the decision of its rival. If firm $i$ is a rational player, at period $t+1$ we have
$$p_i^e(t+1)=\frac{1}{q_i(t+1)+q_{-i}^e(t+1)},$$
where $q_{-i}^e(t+1)$ is the expectation of the output of the rival. Due to the assumption of complete information, which means that $q_{-i}^e(t+1)=q_{-i}(t+1)$, it is acquired that the expected profit of firm $i$ would be
$$\Pi_i^e(t+1)=\frac{q_i(t+1)}{q_i(t+1)+q_{-i}(t+1)}-c_iq_i^2(t+1)$$
The first order condition for profit maximization gives rise to a cubic polynomial equation. To be exact, the condition for the reaction function of firm $i$ would be
\begin{equation}\label{eq:rational-cd-r}
q_{-i}(t+1)-2\,c_iq_i(t+1)(q_i(t+1)+q_{-i}(t+1))^2=0,
\end{equation}
which is simply denoted as $F_i(q_i(t+1),q_{-i}(t+1))= 0$ in the sequel.
The player could maximize its profit by solving the above equation. It is easy to verify that there exists only one positive solution for $q_i(t+1)$ if solving \eqref{eq:rational-cd-r}, but the expression could be quite complex. 
\begin{equation}\label{eq:r_i}
	q_i(t+1)=\frac{\sqrt[3]{2}M}{6c_i}+\frac{\sqrt[3]{4}c_iq_{-i}^2(t+1)}{3M}-\frac{2q_{-i}(t+1)}{3},
\end{equation}
where
$$M=\sqrt[3]{c_i^2	q_{-i}(t+1)(4c_iq_{-i}^2(t+1)+3\sqrt{3}\sqrt{8c_iq_{-i}^2(t+1)+27}+27)}$$

For simplicity, we temporarily denote \eqref{eq:r_i} by 
$$q_i(t+1)=R_i(q_{-i}(t+1)),$$ 
where $R_i$ is called the \emph{best response function} of firm $i$. It is evident that if the two firms in the market are both rational players, the equilibrium (the best decision of output) would be arrived in a shot and there are no dynamics in the system. In order to tackle this problem, Puu introduced the bounded rational player in \cite{Puu1991C}.

A \emph{boundedly rational} player knows the form of the price function, but do not know the rival's decision of the production. If firm $i$ is a boundedly rational player, then it naively expects its competitor to produce the same quantity of output as the last period, i.e., $q_{-i}^e(t+1)=q_{-i}(t)$. Thus,
$$\Pi_i^e(t+1)=\frac{q_i(t+1)}{q_i(t+1)+q_{-i}(t)}-c_iq_i^2(t+1).$$
Then the best response for firm $i$ would be $q_i(t+1)=R_i(q_{-i}(t))$. By now, the following two dynamic duopoly models could be introduced.

\begin{model}[BB]
	We consider a duopoly where two boundedly rational firms simultaneously supply the market. To be exact, we consider a game modeled as
\begin{equation}
	M_{BB}(q_1,q_2): 
	\left\{\begin{split}
		&q_1(t+1)=R_1(q_2(t)),\\
		&q_2(t+1)=R_2(q_1(t)).
	\end{split}
	\right.
\end{equation}
\end{model}

\begin{model}[BR]
	Model BR is similar to Model BB. The only difference is that the second player in the former could know exactly the rival's output at the present period. Thus, the model is described as
\begin{equation}\label{eq:br-map-dim2}
	M_{BR}(q_1,q_2): 
	\left\{\begin{split}
		&q_1(t+1)=R_1(q_2(t)),\\
		&q_2(t+1)=R_2(q_1(t+1)).
	\end{split}
	\right.
\end{equation}

A \emph{local monopolistic approximation} (LMA) player, which even does not know the exact form the price function, is less rational than a boundedly rational player. Specifically, if firm $i$ is an LMA player, then it just can  observe the current market price $p(t)$ and the corresponding total supply $Q(t)$, and is able to correctly estimate the slope $p'(Q(t))$ of the price function around the point $(p(t),Q(t))$.  Then firm $i$ uses such information to conjecture the demand function and expect the price at period $t+1$ to be
$$p_i^e(t+1)=p(Q(t))+p'(Q(t))(Q_i^e(t+1)-Q(t)),$$
where $Q_i^e(t+1)=q_i(t+1)+q_{-i}^e(t+1)$ represents the expected aggregate production of firm $i$ at period $t+1$. Moreover, an LMA player do not know the decision of its rival either, and is assumed to to use the naive expectation, i.e., $q_{-i}^e(t+1)=q_{-i}(t)$. Thus,  
$$p_i^e(t+1)=\frac{1}{Q(t)}-\frac{1}{Q^2(t)}(q_i(t+1)-q_i(t)).$$
The expected profit would be
$$\Pi^e_i(t+1)=q_i(t+1)\left[\frac{1}{Q(t)}-\frac{1}{Q^2(t)}(q_i(t+1)-q_i(t))\right]-c_iq_i^2(t+1).$$
By solving the first order condition, the best response for firm $i$ would be
$$q_i(t+1)=\frac{2\,q_i(t)+q_{-i}(t)}{2(1+c_i(q_i(t)+q_{-i}(t))^2)}.$$
For simplicity, we denote the above map as $q_i(t+1)=S_i(q_i(t),q_{-i}(t))$. 

\begin{model}[LR]
	If replacing the first player in Model BR with an LMA player, we get a new model named LR, which is described as
\begin{equation}
	M_{LR}(q_1,q_2): 
	\left\{\begin{split}
		&q_1(t+1)=S_1(q_1(t),q_2(t)),\\
		&q_2(t+1)=R_2(q_1(t+1)).
	\end{split}
	\right.
\end{equation}
\end{model}

\begin{model}[LB]
Similarly, Model LB is described as

\begin{equation}
	M_{LB}(q_1,q_2): 
	\left\{\begin{split}
		&q_1(t+1)=S_1(q_1(t),q_2(t)),\\
		&q_2(t+1)=R_2(q_1(t)).
	\end{split}
	\right.
\end{equation}	
\end{model}

\begin{model}[LL]
This model is homogeneous and formulated as
\begin{equation}\label{eq:sys-ll}
	M_{LL}(q_1,q_2): 
	\left\{\begin{split}
		&q_1(t+1)=S_1(q_1(t),q_2(t)),\\
		&q_2(t+1)=S_2(q_2(t),q_1(t)).
	\end{split}
	\right.
\end{equation}
\end{model}

Furthermore, we consider a \emph{adaptive} player, which it decides the quantity of production according to its output of the previous period as well as the expectation of its rival. Specifically, if firm $i$ is an adaptive player, then at period $t+1$ this firm naively expects its competitor would produce the same quantity of output as the last period, i.e., $q_{-i}^e(t+1)=q_{-i}(t)$. Then the best response for firm $i$ would be $q_i(t+1)=R_i(q_{-i}(t))$. The adaptive decision mechanism for firm $i$ is that it choose the output $q_i(t+1)$ proportionally to be
$$q_i(t+1)=(1-K)q_i(t)+KR_i(q_{-i}(t)),$$
where $K\in(0,1)$ is a parameter controlling the proportion. It should be notice that an adaptive player degenerate into a boundedly player if we suppose $K=1$. By now, the following two dynamic duopoly models could be introduced.

\begin{model}[AR]
\begin{equation}\label{eq:sys-gr}
	M_{AR}(q_1,q_2): 
	\left\{\begin{split}
		&q_1(t+1)=(1-K)q_1(t)+ KR_1(q_2(t)),\\
		&q_2(t+1)=R_2(q_1(t+1)).
	\end{split}
	\right.
\end{equation}	

\end{model}

\begin{model}[AB]
	\begin{equation}\label{eq:sys-gr}
	M_{AB}(q_1,q_2): 
	\left\{\begin{split}
		&q_1(t+1)=(1-K)q_1(t)+ KR_1(q_2(t)),\\
		&q_2(t+1)=R_2(q_1(t)).
	\end{split}
	\right.
\end{equation}
\end{model}

\begin{model}[AL]
\begin{equation}\label{eq:sys-gl}
	M_{GL}(q_1,q_2): 
	\left\{\begin{split}
		&q_1(t+1)=(1-K)q_1(t)+ KR_1(q_2(t)),\\
		&q_2(t+1)=S_2(q_1(t),q_2(t)).
	\end{split}
	\right.
\end{equation}	
\end{model}

\begin{model}[AA]
	\begin{equation}\label{eq:sys-gg}
	M_{GG}(q_1,q_2): 
	\left\{\begin{split}
		&q_1(t+1)=(1-K_1)q_1(t)+ K_1R_1(q_2(t)),\\
		&q_2(t+1)=(1-K_2)q_2(t)+ K_2R_2(q_1(t)).\\
	\end{split}
	\right.
\end{equation}

\end{model}

\section{Method Description}

The stability analysis of Model LL was first given in \cite{Bischi2007O}, which is restated here.

\begin{proposition}\label{prop:ll}
For Model LL, there exists a unique equilibrium with $q_1,q_2>0$. Moreover, this equilibrium is locally stable for all feasible parameter values, i.e., such parameter values that $c_1,c_2>0$.
\end{proposition}

In \cite{Li2014C}, the first author of this paper proposed an approach of systematically computing semi-algebraic systems, which could also be used in this work to prove similar theoretical results as Proposition \ref{prop:ll}. To illustrate the basic idea of this approach, we reprove the above proposition in a computational style in the sequel.

In order to acquire the equilibria, we set $q_1(t+1)=q_1(t)=q_1$ and $q_2(t+1)=q_2(t)=q_2$ in \eqref{eq:sys-ll}. Moreover, we focus only on such equilibria that $q_1,q_2>0$, then the following system is obtained.
\begin{equation}\label{eq:semi-equi-ll}
	\left\{\begin{split}
		&q_1-S_1(q_1,q_2)=0,\\
		&q_2-S_2(q_1,q_2)=0,\\
		&q_1>0,~q_2>0,\\
		&c_1>0,~c_2>0,
	\end{split}
	\right.
\end{equation}
where $c_1>0$, $c_2>0$ are obvious according to the economic constraints of the parameters. The problem of counting non-vanishing equilibria is equivalent to counting real roots of system \eqref{eq:semi-equi-ll}. Using our approach proposed in \cite{Li2014C}, this problem can be solved in  4 steps as follows.

Step 1. The solutions of the equation part of \eqref{eq:semi-equi-ll} are equivalent to zeros of  
$$P=[\,q_2-c_1q_1(q_1+q_2)^2,~q_1-c_2q_2(q_1+q_2)^2\,]$$
Using the method of triangular decomposition\footnote{The method of triangular decomposition can be viewed as an extension of the method of Gaussian elimination. The main ideas of both are to transform a system into a triangular form. However, the triangular decomposition method  is for polynomial systems, while the Gaussian elimination method  is for linear systems. Refer to \cite{Wu1986B, Li2010D, Kalkbrener1993A, Wang2000C} for more details.}, we decompose $P$ 
into two triangular forms:
$$T_1=[\,q_1,~ q_2\,]$$
and
\begin{equation}
	\begin{split}
T_2=[\,&4\,c_1^3q_1^4-8\,c_1^2c_2q_1^4+4\,c_1c_2^2q_1^4+8\,c_1c_2q_1^2-c_2,\\ &2\,c_1^2q_1^3+2\,c_1c_2q_1^3+4\,c_1c_2q_1^2q_2-c_2q_2\,].		
	\end{split}
\end{equation}
According to the properties of triangular decomposition, the zero set of $P$ is the same as the union of zeros of $T_1$ and $T_2$. Of them, only $T_2$ should be considered as $T_1$ is corresponding to the original equilibrium $(0,0)$. Furthermore, one may observe that the second polynomial in $T_2$ has degree $1$ with respect to $q_2$, thus $q_2$ could be represented as
\begin{equation}\label{eq:ll-q2}
	q_2=\frac{2\,c_1^2q_1^3+2\,c_1c_2q_1^3}{c_2-4\,c_1c_2q_1^2}.
\end{equation}

Step 2. Substitute \eqref{eq:ll-q2} into the inequality $q_2>0$ in \eqref{eq:semi-equi-ll}, we obtain 
$$\frac{2\,c_1^2q_1^3+2\,c_1c_2q_1^3}{c_2-4\,c_1c_2q_1^2}>0,$$
or equivalently
$$(2\,c_1^2q_1^3+2\,c_1c_2q_1^3)(c_2-4\,c_1c_2q_1^2)>0.$$
Thus, system \eqref{eq:semi-equi-ll} is transformed to a univariate system
\begin{equation}\label{eq:ll-univar}
	\left\{\begin{split}
		&4\,c_1^3q_1^4-8\,c_1^2c_2q_1^4+4\,c_1c_2^2q_1^4+8\,c_1c_2q_1^2-c_2=0,\\
		&q_1>0,~(2\,c_1^2q_1^3+2\,c_1c_2q_1^3)(c_2-4\,c_1c_2q_1^2)>0,\\
		&c_1>0,~c_2>0.
	\end{split}
	\right.
\end{equation}

Step 3. Define the border polynomial of system \eqref{eq:ll-univar} to be
$$BP=A_0\cdot\discr(T)\cdot\res(T,q_1)\cdot\res(T,Q),$$
where $\discr(T)$ represents the discriminant of $T$, $\res(T,Q)$ stands for the resultant of $T$ with respect to $Q$, with
\begin{equation}
\begin{split}
	T=&\,4\,c_1^3q_1^4-8\,c_1^2c_2q_1^4+4\,c_1c_2^2q_1^4+8\,c_1c_2q_1^2-c_2,\\
	Q=&\,(2\,c_1^2q_1^3+2\,c_1c_2q_1^3)(c_2-4\,c_1c_2q_1^2),\\
	A_0=&\,4\,c_1^3-8\,c_1^2c_2+4\,c_1c_2^2.
\end{split}
\end{equation}
It is noted that $A_0$ is the leading coefficient of $T$ with respect to $q_1$, i.e., the coefficient of $q_1^4$. We have
\begin{equation}\label{eq:ll-bp-1}
	BP_{LL}=-67108864\,c_1^{11}c_2^{11}(c_1-c_2)^6(c_1+c_2)^{12},
\end{equation}
and its squarefree part is
$$SP_{LL}=c_1c_2(c_1-c_2)(c_1+c_2)$$
It is proved in \cite{Li2014C} that $BP_{LL}=0$, or equivalently $SP_{LL}=0$, divides the parameter space into several connected regions (see Figure \ref{fig:bp1}), and on each of them the number of distinct real solutions is invariant.

\begin{figure}[htbp]
    \centering
    \includegraphics[width=7cm]{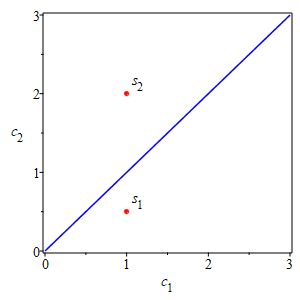}
    \caption{The $2$-dimensional $(c_1,c_2)$ parameter plane of Model LL divided by $SP_{LL}=0$ and the selected sample points.}
    \label{fig:bp1}
\end{figure}

Step 4. From each region of our concerned parameter set with $c_1,c_2>0$, select\footnote{For a simple border polynomial as \eqref{eq:ll-bp-1}, the selection of sample points could be done by hand. Generally, however, the selection could be automated by using, e.g., the partial cylindrical algebraic decomposition (PCAD) method \cite{Collins1991P}.} one sample point:
\begin{equation*}
	s_1=(1, 1/2),~s_2=(1,2).
\end{equation*}
At each sample point, count\footnote{The first thought is to count real solutions by solving the system numerically. However, numerical methods have several shortcomings: 
first, the numerical computation may encounter the problem of instability, which could make the results completely useless; 
second, floating-point errors may cause the problem that it is extremely hard to distinguish between real solutions and complex solutions with tiny imaginary parts;
third, most numerical algorithms only search for a single equilibrium and are nearly infeasible for multiple equilibria. Thus we herein need symbolic methods, e.g., \cite{Xia2002A}, which can be used to count the real solutions exactly.}  the number of distinct real solutions of system \eqref{eq:ll-univar}. We obtain that there exists exactly one real solution for each sample point. This means that system \eqref{eq:ll-univar}, or equivalently \eqref{eq:semi-equi-ll}, has one real solution for any feasible parameter value. Hence, we conclude that the dynamic system \eqref{eq:sys-ll} has one unique equilibrium with $q_1,q_2>0$ for any $c_1,c_2>0$.

It is worth noting that all computations involved in the above 4 steps are symbolic and rigorous. This is to say that our computational approach permit us to acquire analytical results, thus could be used to discover and prove theorems of economic models involving polynomials.

In order to investigate the local stability of this unique equilibrium, the Jacobian matrix
\begin{equation*}
J_{LL}=\left[
	\begin{matrix}
	\partial S_1/\partial q_1 & \partial S_1/\partial q_2\\
	\partial S_2/\partial q_1 & \partial S_2/\partial q_2
	\end{matrix}
\right]
\end{equation*}
plays an ambitious role. We use $\Det(J)$ and $\Tr(J)$ to denote the determinant and the trace of $J$, respectively. According to the Jury's criterion \cite{Jury1976I}, an equilibrium $E$ is locally stable provided that
\begin{equation}\label{eq:ll-jury}
	1+\Tr(J_{LL})+\Det(J_{LL})>0,~1-\Tr(J_{LL})+\Det(J_{LL})>0,~1-\Det(J_{LL})>0.
\end{equation}
Combine these inequalities with system \eqref{eq:semi-equi-ll}, transform the resulting system to a univariate system, and then compute its border polynomial likewise. It is obtained that the squarefree part of the border polynomial is 
$$SP^*_{LL}=c_1c_2(c_1-4\,c_2)(c_1-c_2)(c_1+c_2)(c_1-1/4\,c_2)(c_1^2-7\,c_1c_2+c_2^2).$$
Accordingly, the selected sample points might be
\begin{equation*}
s_1=(2, 1/8),~s_2=(3, 9/16),~s_3=(2, 1),~s_4=(1, 2),~s_5=(9/16, 3),~s_6=(1/8, 2).
\end{equation*}
The parameter plane of Model LL is depicted in Figure \ref{fig:pp-ll}, where $SP^*_{LL}=0$ is represented with blue lines, and the selected sample points are marked in red. It could be verified that the system \eqref{eq:semi-equi-ll}+\eqref{eq:ll-jury} has exactly one real solution at any of these sample points, which means the unique non-vanishing equilibrium of \eqref{eq:sys-ll} is locally stable for all parameter values that satisfy $c_1,c_2>0$.

\begin{figure}[htbp]
    \centering
    \includegraphics[width=7cm]{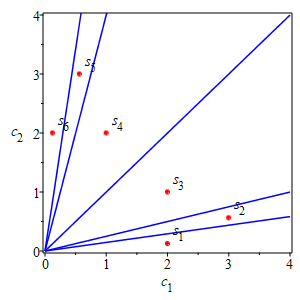}
    \caption{The $2$-dimensional $(c_1,c_2)$ parameter plane of Model LL divided by $SP^*_{LL}=0$ and the selected sample points.}
    \label{fig:pp-ll}
\end{figure}

Indeed, the above proposition has been rigorously proved in Appendix of \cite{Bischi2007O} by explicitly solving the non-vanishing equilibrium. It is lucky that the closed-form solution of the non-vanishing equilibrium exists, which is
\begin{equation}
	\left\{\begin{split}
		q_1^* = \frac{\sqrt{c_2}}{\sqrt{c_1}+\sqrt{c_2}}\frac{1}{\sqrt{2\sqrt{c_1c_2}}},\\
	q_2^* = \frac{\sqrt{c_1}}{\sqrt{c_1}+\sqrt{c_2}}\frac{1}{\sqrt{2\sqrt{c_1c_2}}}.
	\end{split}\right.
\end{equation}
Then substituting the solution into the stability condition by Jury's criterion, it is not hard to verify that all the inequalities in the condition are satisfied. In the next section, it will be found that the verification of the stability through substituting the solution might be impractical for some models. In comparison, our approach do not require to explicitly compute the solution of the considered equilibrium. If the closed-form solution has a super complicated expression or even if there exist no closed-form solutions, our approach could still work.

\section{Stability Analysis}

\subsection{Model LB}

The Jacobian matrix is
\begin{equation}
J_{LB}=\left[
	\begin{matrix}
	\partial S_1/\partial q_1 & \partial S_1/\partial q_2\\
	{\rm d} R_2/{\rm d} q_1 & 0
	\end{matrix}
\right]
\end{equation}
It might be difficult to directly calculate ${\rm d} R_2/{\rm d} q_1$ as the analytical expression of $R_2$ is quite complicated. However, according to \eqref{eq:rational-cd-r}, it is known that
\begin{equation}\label{eq:lb-r2}
	q_1-2\,c_2R_2(q_1)(q_1+R_2(q_1))^2 = 0,
\end{equation}
We could calculate the derivative of the implicit function, if the derivative exists, using the method called implicit differentiation. It is acquired that
$$\frac{{\rm d} R_2}{{\rm d} q_1}=-\frac{4\,c_2q_1q_2+4\,c_2q_2^2-1}{2\,c_2(q_1^2+4\,q_1q_2+3\,q_2^2)}.$$
Hence, the stable equilibria can be described by 
\begin{equation}\label{eq:semi-lb}
	\left\{\begin{split}
		&q_1-S_1(q_1,q_2)=0,\\
		&F_2(q_2,q_1)=0,\\
		&q_1>0,~q_2>0,\\
		&1+\Tr(J_{LB})+\Det(J_{LB})>0,\\
		&1-\Tr(J_{LB})+\Det(J_{LB})>0,\\
		&1-\Det(J_{LB})>0,\\
		&c_1>0,~c_2>0.
	\end{split}
	\right.
\end{equation}

Based on a series of computations, for the above system, we could obtain the squarefree part of the border polynomial
\begin{align*}
	SP^*_{LB}=\,&c_1c_2(c_1+c_2)(-c_2+c_1)(c_1-4\,c_2)(c_1-1/9\,c_2)(c_1^3-28\,c_1^2c_2+4\,c_1c_2^2-c_2^3)\\
	&(c_1^3+4\,c_1^2c_2+12\,c_1c_2^2-c_2^3)(c_1^3+c_1^2c_2+17/4\,c_2^2c_1-1/4\,c_2^3).
\end{align*}
%$$SP^*_{LB}=ca*cb*(ca+cb)*(-cb+ca)*(ca-4*cb)*(ca-1/9*cb)*(ca^3-28*ca^2*cb+4*ca*cb^2-cb^3)*(ca^3+4*ca^2*cb+12*ca*cb^2-cb^3)*(ca^3+ca^2*cb+17/4*cb^2*ca-1/4*cb^3).$$

Accordingly, the sample points might be selected as
$$(1, 1/32),~(1, 1/8),~(1, 1/2),~(1, 2),~(1, 10),~(1, 13),~(1, 18).$$

Counting solutions of system \eqref{eq:semi-lb}, it is known that there exist exactly one real solution at each of the above sample points. Therefore, we finally obtain the following theorem.

\begin{theorem}
For Model LB, there exists a unique equilibrium with $q_1,q_2>0$. Moreover, this equilibrium is locally stable for all feasible parameter values.
\end{theorem}

In comparison, if replacing the cost function in Model LB with a linear function, the case would be different, which was first studied by Cavalli and Naimzada \cite{Cavalli2015Na}. The result is restated as follows.

\begin{proposition}
For Model LB, if replacing the cost functions with $C_1(q_1)=c_1q_1$ and $C_2(q_2)=c_2q_2$, there exists a unique equilibrium with $q_1,q_2>0$. Moreover, this equilibrium is locally stable if $0<c_1/c_2<3+2\sqrt{3}$.
\end{proposition}

\subsection{Model BB}

The Jacobian matrix is
\begin{equation}
J_{BB}=\left[
	\begin{matrix}
	0 & {\rm d} R_1/{\rm d} q_2\\
	{\rm d} R_2/{\rm d} q_1 & 0
	\end{matrix}
\right],
\end{equation}
where ${\rm d} R_1/{\rm d} q_2$ and ${\rm d} R_2/{\rm d} q_1$ could be computed similarly as for Model BL. The stable equilibria can be described by 
\begin{equation}
	\left\{\begin{split}
		&F_1(q_1,q_2)=0,\\
		&F_2(q_2,q_1)=0,\\
		&q_1>0,~q_2>0,\\
		&1+\Tr(J_{BB})+\Det(J_{BB})>0,\\
		&1-\Tr(J_{BB})+\Det(J_{BB})>0,\\
		&1-\Det(J_{BB})>0,\\
		&c_1>0,~c_2>0.
	\end{split}
	\right.
\end{equation}

For the above system, the squarefree part of the polynomial polynomial is
\begin{align*}
	SP^*_{BB}=c_1c_2(c_1-9c_2)(c_1-c_2)(c_1+c_2)(c_1-1/9c_2)(c_1^2-34\,c_1c_2+c_2^2),
\end{align*}
%$$SP^*_{BB}=ca*cb*(ca-9*cb)*(ca-cb)*(ca+cb)*(ca-1/9*cb)*(ca^2-34*ca*cb+cb^2),$$
and the corresponding sample points could be
$$(1, 1/64),~(1, 1/16),~(1, 1/2),~(1, 2),~(1, 10),~(1, 34).$$

We finally obtain the following theorem.

\begin{theorem}
For Model BB, there exists a unique equilibrium with $q_1,q_2>0$. Moreover, this equilibrium is locally stable for all feasible parameter values.
\end{theorem}

The case of linear cost function has been investigated in Puu's seminal work \cite{Puu1991C}. 

\begin{proposition}
For Model BB, if $C_1(q_1)=c_1q_1$ and $C_2(q_2)=c_2q_2$, there exists a unique equilibrium with $q_1,q_2>0$. Moreover, this equilibrium is locally stable if $3-2\sqrt{3}<c_1/c_2<3+2\sqrt{3}$.
\end{proposition}

\subsection{Model BR}

The iteration map \eqref{eq:br-map-dim2} could be transformed into the following one-dimensional system.
\begin{equation}\label{eq:br-map-dim1}
	M_{BR}(q_1): 
	q_1(t+1)=R_1(R_2(q_1(t))).
\end{equation}
\end{model}

The derivative of $q_1(t+1)$ with respect to $q_1(t)$ is critical, which is 
$$\frac{{\rm d} q_1(t+1)}{ {\rm d} q_1(t)} =\frac{{\rm d}[R_1(R_2(q_1(t)))]}{{\rm d} q_1(t)}.$$ 
By the chain rule, we have
\begin{equation}
	\frac{{\rm d}[R_1(R_2(q_1)]}{ {\rm d} q_1 }=\frac{{\rm d} R_1}{ {\rm d} q_2 }\cdot\frac{ {\rm d} R_2}{ {\rm d} q_1} = \frac{4\,c_1q_1q_2+4\,q_1^2-1}{2\,c_1(q_2^2+4\,q_1q_2+3\,q_1^2)}\cdot \frac{4\,c_2q_1q_2+4\,q_2^2-1}{2\,c_2(q_1^2+4\,q_1q_2+3\,q_2^2)}.
\end{equation}
For a one-dimensional dynamic system, an equilibrium $E$ is locally stable provided that at $E$
$$\left|\frac{{\rm d}[R_1(R_2(q_1))]}{ {\rm d} q_1 }\right|<1.$$
Therefore, the stable equilibria are the solutions of the following system.
\begin{equation}
	\left\{\begin{split}
		&F_1(q_1,q_2) = 0,\\
		&F_2(q_2,q_1) = 0,\\
		&q_1>0,~q_2>0,\\
		&1+\frac{{\rm d}[R_1(R_2(q_1))]}{ {\rm d} q_1 }>0,\\
		&1-\frac{{\rm d}[R_1(R_2(q_1))]}{ {\rm d} q_1 }>0,\\
		&c_1>0,~c_2>0.
	\end{split}
	\right.
\end{equation}

The border polynomial is the same as Model BB, then the following theorem is acquired.

\begin{theorem}
For Model BR, there exists a unique equilibrium with $q_1,q_2>0$. Moreover, this equilibrium is locally stable for all feasible parameter values.
\end{theorem}

Moreover, the corresponding model with a linear cost has been investigated in \cite{Cavalli2015Na}.

\begin{proposition}
For Model BR, if $C_1(q_1)=c_1q_1$ and $C_2(q_2)=c_2q_2$, there exists a unique equilibrium with $q_1,q_2>0$. Moreover, this equilibrium is locally stable if $3-2\sqrt{3}<c_1/c_2<3+2\sqrt{3}$.
\end{proposition}

\subsection{Model LR}

Similar as Model BR, this model can be studied by means of a one-dimensional system as follows.
\begin{equation}\label{eq:lr-map-dim1}
	M_{LR}(q_1): 
	q_1(t+1)=S_1(q_1(t),R_2(q_1(t))).
\end{equation}
It is known that 
$$\frac{{\rm d} q_1(t+1)}{ {\rm d} q_1(t)} =\frac{{\rm d}[S_1(q_1(t),R_2(q_1(t)))]}{{\rm d} q_1(t)}.$$ 
By the chain rule, 
\begin{equation}
	\frac{{\rm d}[S_1(q_1,R_2(q_1))]}{{\rm d} q_1} = \frac{\partial S_1}{\partial q_1} + \frac{\partial S_1}{\partial q_2}\cdot \frac{{\rm d} R_2}{{\rm d} q_1}
\end{equation}
Therefore, the stable equilibria are the solutions of the following system.
\begin{equation}
	\left\{\begin{split}
		&q_1-S_1(q_1,q_2)=0,\\
		&F_2(q_2,q_1)=0,\\
		&q_1>0,~q_2>0,\\
		&1+\frac{{\rm d}[S_1(q_1,R_2(q_1))]}{{\rm d} q_1}>0,\\
		&1-\frac{{\rm d}[S_1(q_1,R_2(q_1))]}{{\rm d} q_1}>0,\\
		&c_1>0,~c_2>0.
	\end{split}
	\right.
\end{equation}

The squarefree part of the border polynomial is 
\begin{align*}
	SP^*_{LR}=c_1c_2(c_1+c_2)(-c_2+c_1)(c_1-4\,c_2)(c_1-1/9\,c_2)(c_1^2-21\,c_1c_2+4\,c_2^2),
\end{align*}
%$$SP^*_{LR}=ca*cb*(ca+cb)*(-cb+ca)*(ca-4*cb)*(ca-1/9*cb)*(ca^2-21*ca*cb+4*cb^2),$$
and we could select the following sample points:
$$(1, 1/32),~(1, 1/8),~(1, 1/2),~(1, 2),~(1, 6),~(1, 10).$$

Afterwards, the following theorem is obtained.

\begin{theorem}
For Model LR, there exists a unique equilibrium with $q_1,q_2>0$. Moreover, this equilibrium is locally stable for all feasible parameter values.
\end{theorem}

The case with a linear cost function has also been explored in \cite{Cavalli2015Na}.

\begin{proposition}
For Model LR, if $C_1(q_1)=c_1q_1$ and $C_2(q_2)=c_2q_2$, there exists a unique equilibrium with $q_1,q_2>0$. Moreover, this equilibrium is locally stable if $0<c_1/c_2<7$.
\end{proposition}

\subsection{Model AR}

This model could be equivalently described by
\begin{equation}\label{eq:ar-map-dim1}
	M_{AR}(q_1): 
	q_1(t+1)=(1-K)q_1(t)+ KR_1(R_2(q_1(t))).
\end{equation}
It follows that
$$\frac{{\rm d} q_1(t+1)}{ {\rm d} q_1(t)} =(1-K)+K\cdot\frac{{\rm d} R_1}{ {\rm d} q_2 }\cdot\frac{ {\rm d} R_2}{ {\rm d} q_1}.$$ 
Hence, the stable equilibria are the solutions of the following system.
\begin{equation}
	\left\{\begin{split}
		&F_1(q_1,q_2)=0,\\
		&F_2(q_2,q_1)=0,\\
		&q_1>0,~q_2>0,\\
		&1+ \left((1-K)+K\cdot\frac{{\rm d} R_1}{ {\rm d} q_2 }\cdot\frac{ {\rm d} R_2}{ {\rm d} q_1}\right)>0,\\
		&1- \left((1-K)+K\cdot\frac{{\rm d} R_1}{ {\rm d} q_2 }\cdot\frac{ {\rm d} R_2}{ {\rm d} q_1}\right)>0,\\
		&c_1>0,~c_2>0,~K>0,~1-K>0.
	\end{split}
	\right.
\end{equation}

For the above system, the squarefree part of the border polynomial is
\begin{align*}
	SP^*_{AR}=\,&Kc_1c_2(K-1)(c_1-9\,c_2)(c_1-c_2)(c_1+c_2)(c_1-1/9\,c_2)(c_1^2K^2-2\,c_1c_2K^2\\
	&+c_2^2K^2-3\,c_1^2K+14\,Kc_1c_2-3\,c_2^2K+9/4
	\,c_2^2-41/2\,c_1c_2+9/4\,c_2^2).
\end{align*}
%$$SP^*_{AR}=k*ca*cb*(-1+k)*(ca-9*cb)*(ca-cb)*(ca+cb)*(ca-1/9*cb)*(ca^2*k^2-2*ca*cb*k^2+cb^2*k^2-3*ca^2*k+14*k*ca*cb-3*cb^2*k+9/4*ca^2-41/2*ca*cb+9/4*cb^2).$$
We select the sample points as
$$(1, 1/64, 1/2),~(1, 1/16, 1/2),~(1, 1/16, 3/4),~(1, 1/2, 1/2),~(1, 2, 1/2),~(1, 10, 1/8),~(1, 10, 1/2),~(1, 34, 1/2).$$

The following results are acquired.

\begin{theorem}
For Model AR, there exists a unique equilibrium with $q_1,q_2>0$. Moreover, this equilibrium is locally stable for all feasible parameter values.
\end{theorem}

\begin{theorem}
For Model AR, if $C_1(q_1)=c_1q_1$ and $C_2(q_2)=c_2q_2$, there exists a unique equilibrium with $q_1,q_2>0$. Moreover, this equilibrium is locally stable if 
$$c_1^2K+2\,c_1c_2K+c_2^2K-8\,c_1c_2<0.$$
%$$ca^2*k+2*ca*cb*k+cb^2*k-8*ca*cb<0.$$
\end{theorem}

%
%\begin{figure}[htbp]
%  \centering
%    \subfigure[]{\includegraphics[width=0.4\textwidth]{fig/gr-1.png}} 
%    \subfigure[]{\includegraphics[width=0.4\textwidth]{fig/gr-2.png}}\\
%    \subfigure[$c_1=c_2$.]{\includegraphics[width=0.4\textwidth]{fig/gr-2d-1.png}} 
%    \subfigure[$K=1$.]{\includegraphics[width=0.4\textwidth]{fig/gr-2d-2.png}}
%  \caption{The 3-dimensional $(c_1,c_2,K)$ parameter space of Model GR. The red surface is $R_{GR}^1=0$, and the blue surface is $R_{GR}^2=0$.}
%    \label{fig:par-gr}
%\end{figure}
%
%
%
%\begin{proposition}
%	For Model GR, if $c_1>4$ or $c_2>3$, the stable region for the linear costs $C_i(q_i)=c_iq_i$ is strictly contained in that for the quadratic costs $C_i(q_i)=c_iq_i^2$.
%\end{proposition}

\subsection{Model AB}

The Jacobian matrix is
\begin{equation}
J_{AB}=\left[
	\begin{matrix}
	1-K & K\cdot {\rm d} R_1/{\rm d} q_2\\
	{\rm d} R_2/{\rm d} q_1 & 0
	\end{matrix}
\right]
\end{equation}
Hence, the stable equilibria can be described by 
\begin{equation}\label{eq:semi-lb}
	\left\{\begin{split}
		&F_1(q_1,q_2)=0,\\
		&F_2(q_2,q_1)=0,\\
		&q_1>0,~q_2>0,\\
		&1+\Tr(J_{AB})+\Det(J_{AB})>0,\\
		&1-\Tr(J_{AB})+\Det(J_{AB})>0,\\
		&1-\Det(J_{AB})>0,\\
		&c_1>0,~c_2>0,~K>0.
	\end{split}
	\right.
\end{equation}

%SP= k*ca*cb*(-1+k)*(ca+cb)*(ca-cb)*(ca-9*cb)*(ca-1/9*cb)*(ca^2*k^2-2*ca*cb*k^2+cb^2*k^2+6*ca^2*k+52*ca*cb*k+6*cb^2*k+9*ca^2-82*ca*cb+9*cb^2)*(ca^2*k^2-2*ca*cb*k^2+cb^2*k^2-6*ca^2*k-52*ca*cb*k-6*cb^2*k+9*ca^2-82*ca*cb+9*cb^2)*(ca^2*k^2-34*ca*cb*k^2+cb^2*k^2-6*ca^2*k+108*ca*cb*k-6*cb^2*k+9*ca^2-82*ca*cb+9*cb^2)
%
%[[[1, 1/128, 1/2], 1], [[1, 1/64, 1/2], 1], [[1, 1/32, 1/2], 1], [[1, 1/32, 31/32], 1], [[1, 1/16, 1/4], 1], [[1, 1/16, 1/2], 1], [[1, 1/2,
%1/2], 1], [[1, 1/2, 29/32], 1], [[1, 1/2, 15/16], 1], [[1, 2, 1/2], 1], [[1, 2, 29/32], 1], [[1, 2, 15/16], 1], [[1, 10, 1/16], 1], [[1, 10, 1/2], 1], [[1, 26, 1/2], 1], [[1, 26,
%7/8], 1], [[1, 34, 1/2], 1], [[1, 98, 1/2], 1]], [[[1, 1/128, 1/2], 1], [[1, 1/64, 1/2], 1], [[1, 1/32, 1/2], 1], [[1, 1/32, 31/32], 1], [[1, 1/16, 1/4], 1], [[1, 1/16, 1/2], 1],
%[[1, 1/2, 1/2], 1], [[1, 1/2, 29/32], 1], [[1, 1/2, 15/16], 1], [[1, 2, 1/2], 1], [[1, 2, 29/32], 1], [[1, 2, 15/16], 1], [[1, 10, 1/16], 1], [[1, 10, 1/2], 1], [[1, 26, 1/2], 1],
%[[1, 26, 7/8], 1], [[1, 34, 1/2], 1], [[1, 98, 1/2], 1]]

Similarly, we obtain the following results.

\begin{theorem}
For Model AB, there exists a unique equilibrium with $q_1,q_2>0$. Moreover, this equilibrium is locally stable for all feasible parameter values.
\end{theorem}

\begin{theorem}
For Model GB, if $C_1(q_1)=c_1q_1$ and $C_2(q_2)=c_2q_2$, there exists a unique equilibrium with $q_1,q_2>0$. Moreover, this equilibrium is locally stable if 
$$c_1^2K-2\,c_1c_2K+c_2^2K-4\,c_1c_2<0.$$
%$$ca^2*k-2*ca*cb*k+cb^2*k-4*ca*cb<0.$$
\end{theorem}

\subsection{Model AL}

The Jacobian matrix is
\begin{equation}
J_{AL}=\left[
	\begin{matrix}
1-K & K\cdot {\rm d} R_1/{\rm d} q_2\\
	\partial S_2/\partial q_1 & \partial S_2/\partial q_2
	\end{matrix}
\right]
\end{equation}
Hence, the stable equilibria can be described by 
\begin{equation}\label{eq:semi-gl}
	\left\{\begin{split}
		&F_1(q_1,q_2)=0,\\
		&S_2(q_2,q_1)=0,\\
		&q_1>0,~q_2>0,\\
		&1+\Tr(J_{AL})+\Det(J_{AL})>0,\\
		&1-\Tr(J_{AL})+\Det(J_{AL})>0,\\
		&1-\Det(J_{AL})>0,\\
		&c_1>0,~c_2>0,~K>0.
	\end{split}
	\right.
\end{equation}

As a consequece, we obtain the following results.

\begin{theorem}
For Model AL, there exists a unique equilibrium with $q_1,q_2>0$. Moreover, this equilibrium is locally stable for all feasible parameter values.
\end{theorem}

\begin{theorem}
For Model AL, if $C_1(q_1)=c_1q_1$ and $C_2(q_2)=c_2q_2$, there exists a unique equilibrium with $q_1,q_2>0$. Moreover, this equilibrium is locally stable if
$$3\,c_1^2K+2\,c_1c_2K-c_2^2K+4\,c_1c_2<0.$$
%$$3*ca^2*k+2*ca*cb*k-cb^2*k+4*ca*cb<0.$$
\end{theorem}

\subsection{Model AA}

The Jacobian matrix is
\begin{equation}
J_{AA}=\left[
	\begin{matrix}
	1-K_1 & K_1\cdot {\rm d} R_1/{\rm d} q_2\\
	K_2\cdot {\rm d} R_2/{\rm d} q_1 & 1-K_2
	\end{matrix}
\right]
\end{equation}
Hence, the stable equilibria can be described by 
\begin{equation}\label{eq:semi-gg}
	\left\{\begin{split}
		&F_1(q_1,q_2)=0,\\
		&F_2(q_2,q_1)=0,\\
		&q_1>0,~q_2>0,\\
		&1+\Tr(J_{AA})+\Det(J_{AA})>0,\\
		&1-\Tr(J_{AA})+\Det(J_{AA})>0,\\
		&1-\Det(J_{AA})>0,\\
		&c_1>0,~c_2>0,~K_1>0,~K_2>0.
	\end{split}
	\right.
\end{equation}

The following results are acquired.

\begin{theorem}
For Model AA, there exists a unique equilibrium with $q_1,q_2>0$. Moreover, this equilibrium is locally stable for all feasible parameter values.
\end{theorem}

The following proposition is a known result first addressed by Agliari \cite{Agliari2006H}.
 
\begin{proposition}
For Model AA, if $C_1(q_1)=c_1q_1$ and $C_2(q_2)=c_2q_2$, there exists a unique equilibrium with $q_1,q_2>0$. Moreover, this equilibrium is locally stable if 
$$c_1^2K_1K_2+2\,c_1c_2K_1K_2+c_2^2K_1K_2-4\,c_1c_2K_1-4\,c_1c_2K_2<0.$$
%$$ca^2*k*l+2*ca*cb*k*l+cb^2*k*l-4*ca*cb*k-4*ca*cb*l<0.$$
\end{proposition}

%\section{Bifurcation Analysis}

%\section{Concluding Remarks}
%
%Fisher \cite{Fisher1961T}

%\section*{Acknowledgements}
%
%The authors are grateful to the anonymous referees for their helpful comments. This work has been supported by Philosophy and Social Science Foundation of Guangdong under Grant GD21CLJ01, Major Research and Cultivation Project of Dongguan City College under Grant 2021YZDYB04Z and Social Development Science and Technology Project of Dongguan under Grant 20211800900692. 

\bibliographystyle{abbrv}
\bibliography{duopoly.bib}

%\section*{Appendix}
%\tiny
%\begin{align*}
%
%\end{autobreak}\\
%
%\begin{autobreak}

%\end{align*}
%
%
%\newpage
%
%\section*{Highlights}
%
%\begin{enumerate}
%    \item 
%    
%    \item
%    
%    \item
%    
%\end{enumerate}
%
%
%
%\section*{Cover Letter}
%\begin{lstlisting}[breaklines=true, columns=flexible]
%
%Dear editor,
%
%I would like to submit the enclosed manuscript entitled "***" by *** for possible publication in Communications in ***. 
%
%[abstract]
%
%Highlights of this work include the following.
%
%[highlights]
%
%The contact information of the corresponding author is as follows.
%
%   Name: ***;
%   Address: ***; 
%   E-mail: ***;
%   Mobile: ***;
%   Fax: N/A.
%
%Thank you very much for consideration!
%
%Sincerely yours,
%***
%
%\end{lstlisting}

\end{CJK}
\end{document}